\documentclass{article} 
\usepackage{iclr2019_conference,times}
\usepackage{latexsym, epsfig, verbatim,  lscape, color}
\usepackage{epsfig,booktabs}
\usepackage{graphicx}
\usepackage{caption}
\usepackage{subcaption}
\usepackage{bold-extra}
\usepackage{amsfonts}
\usepackage{amsmath, amssymb,caption}
\usepackage{algorithm}
\newcommand{\indep}{\raisebox{0.05em}{\rotatebox[origin=c]{90}{$\models$}}}
\usepackage{url}
\usepackage{algpseudocode}
\newtheorem{proposition}{Proposition}
\newtheorem{thm}{Theorem}[section]

\newtheorem{definition}[thm]{Definition}
\usepackage{url}
\usepackage{algpseudocode}
\newcommand\ci{\perp\!\!\!\perp}

\usepackage{amsmath,amsfonts,bm}

\def\eqref#1{equation~\ref{#1}}

\def\1{\bm{1}}

\def\vx{{\bm{x}}}

\def\mX{{\bm{X}}}

\DeclareMathAlphabet{\mathsfit}{\encodingdefault}{\sfdefault}{m}{sl}
\SetMathAlphabet{\mathsfit}{bold}{\encodingdefault}{\sfdefault}{bx}{n}

\usepackage{hyperref}
\usepackage{url}

\title{Auto-Encoding Knockoff Generator for FDR 
Controlled Variable Selection}

\author{Ying Liu  \\
Division of Biostatistics\\
Medical College of Wisconsin\\
Milwaukee, WI  \\
\texttt{summeryingl@gmail.com} \\
\AND
Cheng Zheng \\
Joseph J. Zilber School of Public Health \\
University of Wisconsin-Milwaukee \\
Milwaukee, WI \\
\texttt{zhengc@uwm.edu}
}

\iclrfinalcopy 
\begin{document}

\maketitle

\begin{abstract}
A new statistical procedure (Model-X \cite{candes2018}) has provided a way to identify important factors using any supervised learning method controlling for FDR. This line of research has shown great potential to expand the horizon of machine learning methods beyond the task of prediction, to serve the broader needs in scientific researches for interpretable findings. However, the lack of a practical and flexible method to generate knockoffs remains the major obstacle for wide application of Model-X procedure. This paper fills in the gap by proposing a model-free knockoff generator which approximates the correlation structure between features through latent variable representation. We demonstrate our proposed method can achieve FDR control and better power than two existing methods in various simulated settings and a real data example for finding mutations associated with drug resistance in HIV-1 patients.
\end{abstract}

\section{Introduction}
In the past decades, the machine learning methods have achieved great advancement in improving prediction accuracy. However, prediction accuracy is not the sole quest of 'big data' analysis. A universal aim across a lot of scientific disciplines is to identify the causes of certain outcome. For example, in new drug development, a personalized medicine strategy can be formed by identifying the gene markers that is biological linked to certain drug response or resistance. In electronic health record analysis, the goal is to identify 'actionable' factors for the purpose of reducing cost and improving the quality of care. In epidemiology or sociology studies, it is of interest to identify protection or risk factors, especially those that can be changed through public policy or civil planning to improve public welfare. In these scientific endeavors, the need is to identify the important factors associated certain outcome, so that further confirmatory investigation (e.g. randomized international studies) could be conducted to expand knowledge or change future actions for a better outcome. For this purpose, we are interested in procedures that control the Type I error, which is the chance of a false discovery. With a large amount of factors to test, controlling the chance of any false discovery (family-wise error) is too stringent. Therefore the objective is usually relaxed to control the False Discovery Rate (FDR) \citep{BH1995}.

A recent breakthrough in the statistical theory, i.e. the Model-X framework\citep{barber2015,candes2018}, provided a general solution. It can be incorporated with any machine learning models to select true signals associated with the outcome, with rigorous control for FDR. This line of research showed light on expanding the horizon of supervised learning methods beyond prediction. However, the implementation of Model-X requires the generation of the so called 'knockoffs', which has very limited existing methods. The goal of our paper is to fill in the gap by proposing a model-free method for generating knockoffs that is suitable for any data type. It can also be efficiently implemented leveraging on the power of of the recent development in deep generative models. Consequently, with the creation of such a generator, we equipped every supervised learning method the ability to conduct FDR controlled variable selection.

We consider the following problem \citep{candes2018}. There are $N$ i.i.d. samples $(X^{(i)},Y^{i})$ from a population, where the predictors are random variables $X=(X_1,\dots,X_p)$. The outcome $Y$ only depend on a subset of predictors $S\in \{1,2,\dots,p\}$, i.e. given $\{X_j, j\in S\}$, $Y$ is independent of the other $X_j$'s. The smallest set $S$ that satisfy this requirement is considered as true factors. The goal is to find a good estimator $\hat{S}$ for $S$ with control of the $FDR =E\Big(\frac{\#\{j:j\in \hat{S}\backslash S\}}{\#\{j:j\in\hat{S}\} \vee 1}\Big)$. 
 \cite{barber2015} first proposed the method for linear regression with Gaussian errors, where the design matrix $\mX$ is assumed to be fixed. \cite{candes2018} greatly generalized it to the Model-X framework. 
In contrast to the traditional statistical models focusing on the conditional distribution $Y|X_1, \dots, X_p$, Model-X makes no assumption for the conditional distribution, but assume $X$ has a known distribution for the purpose of generating knockoffs $\tilde{X}$ that satisfies Definition \ref{knockoffdef}.
\begin{definition} 
{\bf Model-X knockoffs} for the family of random variables $X=(X_1, \dots, X_p)$ are a new family of random variables $\tilde{X}=(\tilde{X}_1,\dots,\tilde{X}_p)$ constructed with the following two properties: (1) for any subset $S \subset \{1,\dots,p\}$, 
$(X,\tilde{X})_{\textit{swap}(S)} =_{d} (X,\tilde{X});$ 
(2) $\tilde{X} \ci Y | X$ if there is a response Y. (2) is guaranteed if $\tilde{X}$ is constructed without looking at Y.
\label{knockoffdef}
\end{definition}
\begin{proposition}[Cand\`{e}s 2018] 
 The random variables $\tilde{X}=(\tilde{X}_1,\dots,\tilde{X}_p)$  are model-X knockoffs for $X=(X_1, \dots, X_p)$\\ if and only if  $(X_j,X_{-j},\tilde{X}_j,\tilde{X}_{-j})=_d(\tilde{X}_j,X_{-j},X_j,\tilde{X}_{-j})$, for any $j\in\{1,\dots,p\}$. And $Y\indep\tilde{X} |X$.
 \label{pair} 
\end{proposition}
To implement Model-X with any machine learning models, one need to define the model-specific feature statistics $W_j=w_j([X,\tilde{X}],y)$ that has the flip-sign property, $$W_j([X,\tilde{X}]_{\textit{swap}(S)},y) = (1-2 \1_\mathrm{\{j\in S\}}) W_j([X,\tilde{X}],y).$$  According to Thm 3.4 in \cite{candes2018},  the following procedure (Knockoff) can estimate the $\hat{S}$ controlling for modified FDR, i.e. $E\Big(\frac{\#\{j:j\in \hat{S}\backslash S\}}{\#\{j:j\in\hat{S}\} +q^{-1}}\Big) \le q$.  First, compute the threshold  $\tau=\min\big\{t>0: \frac{\#\{j:W_j\le-t\}}{\#\{j:W_j\ge t\}}\le q\big\},$ then the selected set is
\begin{equation}
\textrm{(Knockoff):  } \hat{S}=\{j:W_j\ge \tau\}.
\label{knockoff}
\end{equation}
More conservatively, the following procedure (Knockoff+) can control for the FDR, choose a threshold $\tau+=\min\big\{t>0: \frac{1+\#\{j:W_j\le-t\}}{\#\{j:W_j\ge t\}}\le q\big\},$ and select the variables in the set 
\begin{equation}
\textrm{(Knockoff+):  } \hat{S}=\{j:W_j\ge \tau+\}.
\label{knockoff+}
\end{equation}
  An example of the feature statistics is the signed max lambda statistics in $L_1$ penalized regressions \cite{barber2015}: for the concatenated design matrix $[X,\tilde{X}]$, define $\hat{\beta}(\lambda)=\arg\min_{b}\Big\{Loss(y,[X \tilde{X}]b)+\lambda\|b\|_{L1}\Big\}$.
For each feature $X_j$ and its knockoff $\tilde{X_j}$, let $Z_j= \sup\{\lambda: \hat{\beta}_{2j-1}(\lambda)\neq 0\}$ and $\tilde{Z_j}= \sup\{\lambda: \hat{\beta}_{2j}(\lambda)\neq 0\}$. The signed lambda statistics is defined as $W_j= (Z_j\vee \tilde{Z_j}) \textrm{sign}(Z_j-\tilde{Z_j})$. Various feature statistics has been proposed for common learning methods \citep{candes2018,gimenez2018}, however the current knockoff generation methods are still limited to make Model-X generally applicable on any real data set.

Although \cite{candes2018} provided a general algorithm for generating Model-X knockoffs, i.e. the Sequential Conditional Independent Pairs algorithm, it needs to sample from the conditional distribution of $\tilde{X}_j$ from $f(X_j|X_{-j},\tilde{X}_{1:j-1})$ which becomes intractable with slightly complex distributional assumption for $X$. \cite{candes2018} also proposed a second-order approach by matching first two moments of $(X,\tilde{X})_{\textit{swap}(S)}$ and $(X,\tilde{X})$, which satisfies Def.~\ref{knockoffdef} when $X$ is from Gaussian distribution. \cite{sesia2018} proposed an algorithm to sample knockoffs when $X$ is from Hidden Markov Model. \cite{gimenez2018} proposed algorithm that applicable for $X$ from simple Bayesian network models (such as mixture Gaussian) with distributional assumptions for certain conditional probabilities which need to be estimated and sampled from. 

In this paper, we relaxes the distributional assumptions largely from all the existing methods. Our procedure assumes there is a (multivariate) latent variable $Z$, conditional on which $X$'s are mutually independent. We propose a procedure to generate $X$'s knockoff from conditional distributions of $Z|X$ and $X|Z$. And we also provide a FDR bound when these estimated conditional probabilities is compatible with the distribution of $X$. In practice, we assume $Z$ and the conditional distributions are from parametric families, which is adaptive to different real data application, and the parameters can be approximated by deep neural networks. In this manner, the algorithms developed in the line of research on variational autoencoder, provide a flexible and practical tool for implementing our proposed Knockoff generation method. In the next session, we provide the formal derivation and justification for this approach.

\section{A Model-Free Knockoff Generator With Latent Encoding Variables.}
The random vector of covariates $X=(X_1,X_2,\dots,X_p)$ are from $p_X(x)$ (X can be continuous, discrete or mixed ).  There exists a vector of continuous latent random variables $Z$ from distribution $p_Z(z)$. 
Each component of the covariates $ X_j \sim p_\epsilon (f_i(Z)), j =1,\dots,p$, such that $\epsilon_j$'s are mutually independent given $Z$.  For example, when the predictor is continuous, $X_j\sim \mathcal{N}(f_j(Z),\sigma^2_j)$; and when the predictor is binary, $X_j \sim \textrm{Bernoulli}(f_j(Z))$.  
Given the observed samples $\{\vx^{(i)}\}_{i=1}^N$ from $p_X(x)$, to estimate $p_Z(z)$ and $p_{X|Z}$ may not be an identifiable problem. However, we just need to find one $Z$ that approximately satisfy the above condition in practice for our purpose of knockoff generation. 
Then we propose to generate knockoff $\tilde{X}$ as Algorithm~\ref{VAE}.
\begin{algorithm}
\caption{Auto-Encoding Knockoff Generator.}
\begin{algorithmic}[1]
\State Assume two working models from parametric families: a) an encoder $Q_{Z|X}(z|x;\theta)$ to approximate $p_{Z|X}(z|x)$; b) a decoder $Q_{X|Z}(x|z;f,q_\epsilon)$ to approximate $p_{X|Z}(x|z)$.
\State Train model parameters to get estimates $\hat{\theta},\hat{f}$. And let $q_\epsilon$ be the working noise distribution to generate $\tilde{X}$ given $\hat{f}(\tilde{Z})$, and $\epsilon$ is element-wise independent $q_\epsilon=\prod_{j} q_{\epsilon_j}$.
\State Generate $\tilde{Z}$ from $Q_{Z|X}(z|X;\hat{\theta})$. 
\State Generate $\tilde{X}$ from $Q_{X|Z}(x|\tilde{Z};\hat{f},q_\epsilon)$.
\end{algorithmic}
\label{VAE}
\end{algorithm}

Algorithm \ref{VAE} can be efficiently implemented with but not limited to the existing algorithms developed in the line of research on Variational Autoencoder (VAE) \citep{kingma2014,rezende2014,maddison2017,eric2017}.
For example, in the common VAE implementation, $Z$ is assumed tobe from its prior $\mathcal{N}(0,I_q)$, the working model $Q_{Z_i|X}$ is assumed to be from $\mathcal{N}(\mu_i(X),\sigma^2_i(X))$, and $Q_{X_i|Z}$ is $\mathcal{N}(f_i(Z),\sigma^2)$, where $\mu_i(\bullet)$ and $\sigma^2_i(\bullet)$ and $f(\bullet)$ are functions to be approximated by neural networks. The $q_\epsilon$ in this case is i.i.d. normal distribution with infinitesimal variance. So in practice $\tilde{X}$ is generated as $\hat{f}(\tilde{Z})$. Another example is to approximate discrete  $Z$'s by the concrete distribution \citep{eric2017,maddison2017}. In this case $\tilde{X}$ is generated from $Bernoulli(\hat{f}(\tilde{Z}))$ and $\epsilon=\tilde{X}-\hat{f}(\tilde{Z})$ are element-wise independent.  These two methods provide efficient algorithms for implementing Algorithm 1 in specific cases. However Algorithm 1 and the following justification are more general. We do not assume $Z$'s from independent prior distribution, and $Z|X$ do not need to be element-wise independent either. See more discussions in Section 5 paragraph 2. 

Notice that the Step 3 in Algorithm 1 is different from the deep generative model, the latter generate new image by first generating a $Z$ from the predetermined prior distribution $p_Z(z)$. However, in step 3 of our algorithm, the new $\tilde{Z}$ is generated from the conditional distribution $Q_{Z|X}$. The following Theorem 2.1 justifies why we propose procedure as so.

\begin{thm} 
For any vector of random variables $Z$, such that conditional on $Z$, $X_i$'s are mutually independent. If $\tilde {Z} $ is from $p_{Z|X}(\bullet|X)$, then $\tilde {X} $ from $p_{X|Z}(\bullet|\tilde{Z})$ is model-X knockoffs.
\end{thm}
\textbf{Proof:} 
For any $j\in{1,\dots, p}$ and sets $A_j$, $B_j$, $A_{-j}$, $B_{-j}$, 
\begin{eqnarray*}
&&P(X_j\in A_j,\tilde{X}_j\in B_j,X_{-j}\in A_{-j},\tilde{X}_{-j}\in B_{-j})
=\int p_{X|Z}(A_j,A_{-j}|Z)p_{X|Z}(B_j,B_{-j}|Z)dF(Z)\\
&=&\int p_{X_j|Z}(A_j|Z)p_{X_{-j}|Z}(A_{-j}|Z)p_{X_j|Z}(B_j|Z)p_{X_{-j}|Z}(B_{-j}|Z)dF(Z)\\
&=&\int p_{X_j|Z}(B_j|Z)p_{X_{-j}|Z}(A_{-j}|Z)p_{X_j|Z}(A_j|Z)p_{X_{-j}|Z}(B_{-j}|Z)dF(Z)\\
&=&\int p_{X|Z}(B_j,A_{-j}|Z)p_{X|Z}(A_j,B_{-j}|Z)dF(Z)
=P(X_j\in B_j,\tilde{X}_j\in A_j,X_{-j}\in A_{-j},\tilde{X}_{-j}\in B_{-j})\\
\end{eqnarray*}
We have shown $X_j$ and $\tilde{X}_j$ are exchangeable in the joint distribution, then $\tilde{X}$ is the Model-X knockoff by Proposition 1.



To provide the FDR bound for the knockoffs generated from the estimated encoder and decoder, we do not assume there exists a $Z$ satisfies the condition of Theorem 2.1 exactly. Instead, we assume the working models and observed distribution of $X$ are approximately compatible as in the following theorem. We then apply the results per \cite{barber2018} to show FDR control, where they showed the FDR derived from approximate knockoffs
is bounded by a function of the observed KL divergence between $(X_j,\tilde{X}_j,X_{-j},\tilde{X}_{-j})$ and $(\tilde{X}_j,X_j,X_{-j},\tilde{X}_{-j})$ as below:
\begin{eqnarray*}
\widehat{KL}_j&=&\log \left(\frac{P(X_j,\tilde{X}_j,X_{-j},\tilde{X}_{-j})}{P(\tilde{X}_j,X_j,X_{-j},\tilde{X}_{-j})}\right)=\sum_{i=1}^n \log \left(\frac{P(X_{ij},\tilde{X}_{ij},X_{i,-j},\tilde{X}_{i,-j})}{P(\tilde{X}_{ij},X_{ij},X_{i,-j},\tilde{X}_{i,-j})}\right)
\end{eqnarray*}

\begin{thm} 
Assume the observed marginal distribution $p_X(x)$ and the working models ($Q_{Z|X}$, $Q_{X|Z}$, $q_\epsilon$ estimated in Algorithm 1) are approximately compatible as following: there exists a random vector $Z$ from certain distribution $p_Z(z)$ and $a_n\rightarrow 0$, such that $$\sup_x|\log \left(\frac{q_{\epsilon}(x)}{p_{\hat{\epsilon}}(x)}\right)|\leq a_n \textrm{ and } \sup_{z,x}|\log \left(\frac{p_{Z|\hat{X}}(z|x)}{ Q_{Z|X}(z|x)}\right)|\leq a_n.$$
Where the density of $\hat{\epsilon}=X-\hat{f}(Z)$ is denoted as $p_{\hat{\epsilon}}$, considering $\hat{f}$ as a fixed function. And denote $\hat{X}=\hat{f}(Z)+\epsilon$ as a random variable generated from the decoder $Q_{X|Z}(x|z=Z;\hat{f},q_{\epsilon})$.

 Then the FDR can be controlled at $q\exp\{8na_n^2+8\sqrt{n\log(p)}a_n\}$.
\end{thm}
\textbf{Proof:}  Consider $\tilde{\hat{X}}$ which is another independent sample from $Q_{X|Z}(x|z=Z;q_\epsilon)$ when sampling $\hat{X}$ using the same $Z$. 
Since $q_\epsilon$ is element-wise independent, conditions in Theorem 2.1 holds, thus $\tilde{\hat{X}}$ is a model-X knockoff of $\hat{X}$ and

\begin{eqnarray*}
\log \left(\frac{P(\hat{X}_{j}\in A_j,\tilde{\hat{X}}_{j}\in B_j,\hat{X}_{-j}\in A_{-j},\tilde{\hat{X}}_{-j}\in B_{-j})}{P(\hat{X}_{j}\in B_j,\tilde{\hat{X}}_{j}\in A_j,\hat{X}_{-j}\in A_{-j},\tilde{\hat{X}}_{-j}\in B_{-j})}\right)=0.
\end{eqnarray*}
Notice that by our assumptions
\begin{eqnarray*}
&&\sup_x|\log \left(\frac{p_{\hat{X}}(x)}{p_{X}(x)}\right)|
\leq \sup_{x,z}|\log \left(\frac{p_{\hat{X}|Z}(x|z)}{p_{X|Z}(x|z)}\right)|
= \sup_{x,z}|\log \left(\frac{q_{\epsilon}(x-\hat{f}(z))}{p_{\hat{\epsilon}}(x-\hat{f}(z))}\right)|\leq a_n
\end{eqnarray*}


\begin{eqnarray*}
\sup_{x,\tilde{x}}|\log \left(\frac{p_{\tilde{\hat{X}}|\tilde{X}}(\tilde{x}|x)}{p_{\tilde{X}|X}(\tilde{x}|x)}\right)|
=\sup_{x,z,\tilde{x}}|\log \left(\frac{\int q_{\epsilon}(\tilde{x}-\hat{f}(z))p_{Z|\hat{X}}(z|x)dz}{\int q_{\epsilon}(\tilde{x}-\hat{f}(z))Q_{Z|X}(z|x)dz}\right)|\leq \sup_{z,x} |\log \left(\frac{p_{Z|\hat{X}}(z|x)}{ Q_{Z|X}(z|x)}\right)|\leq a_n
\end{eqnarray*}

So $\sup_{x,\tilde{x}}|\log \left(\frac{p_{(\hat{X},\tilde{\hat{X}})}(x,\tilde{x})}{p_{(X,\tilde{X})}(x,\tilde{x})}\right)|\leq 2a_n$  and we can bound the log likelihood ratio by 
{\small
\begin{eqnarray*}
&&\log \left(\frac{P(X_{j}\in A_j,\tilde{X}_{j}\in B_j,X_{-j}\in A_{-j},\tilde{X}_{-j}\in B_{-j})}{P(X_{j}\in B_j,\tilde{X}_{j}\in A_j,X_{-j}\in A_{-j},\tilde{X}_{-j}\in B_{-j})}\right)\\
&=&\log \left(\frac{P(\hat{X}_{j}\in A_j,\tilde{\hat{X}}_{j}\in B_j,\hat{X}_{-j}\in A_{-j},\tilde{\hat{X}}_{-j}\in B_{-j})}{P(\hat{X}_{j}\in B_j,\tilde{\hat{X}}_{j}\in A_j,\hat{X}_{-j}\in A_{-j},\tilde{\hat{X}}_{-j}\in B_{-j})}\right)
+\log \left(\frac{P(X_{j}\in A_j,\tilde{X}_{j}\in B_j,X_{-j}\in A_{-j},\tilde{X}_{-j}\in B_{-j})}{P(\hat{X}_{j}\in A_j,\tilde{\hat{X}}_{j}\in B_j,\hat{X}_{-j}\in A_{-j},\tilde{\hat{X}}_{-j}\in B_{-j})}\right)\\
&&+\log \left(\frac{P(\hat{X}_{j}\in B_j,\tilde{\hat{X}}_{j}\in A_j,\hat{X}_{-j}\in A_{-j},\tilde{\hat{X}}_{-j}\in B_{-j}))}{P(X_{j}\in B_j,\tilde{X}_{j}\in A_j,X_{-j}\in A_{-j},\tilde{X}_{-j}\in B_{-j})}\right)
\leq 0+2a_n+2a_n=4a_n
\end{eqnarray*}}
So we can bound 
\begin{eqnarray*}
\max_{j=1,\cdots,p}\widehat{KL}_j\leq 8na_n^2+8\sqrt{n\log(p)}a_n.
\end{eqnarray*}
And FDR has the bound in the statement by applying Lemma 2 in \cite{barber2018}.
Specifically, if $a_n=o((n\log (p))^{-1/2})$, then FDR will be asymptotically bounded by $q+o(1)$.

\section{Simulation}
\begin{figure}[h]
\begin{subfigure}[b]{.45\linewidth} 
\includegraphics[width=0.9\textwidth]{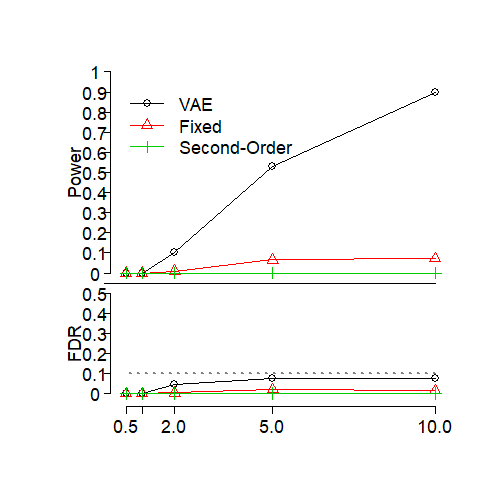}
\caption{Gaussian $m=10$, FDR$=0.1$}
\end{subfigure}\begin{subfigure}[b]{.45\linewidth} 
\includegraphics[width=0.9\textwidth]{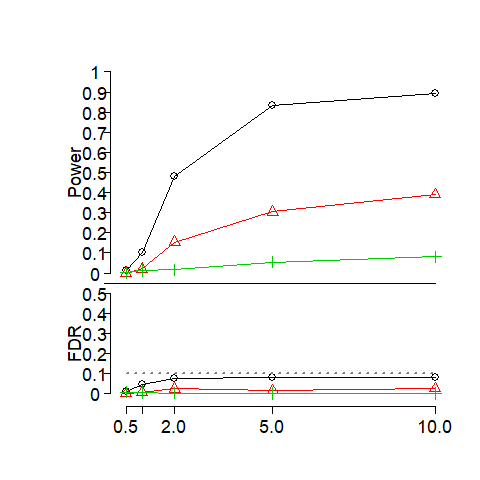}
\caption{Gaussian $m=20$, FDR$=0.1$}
\end{subfigure}
\begin{subfigure}[b]{.45\linewidth} 
\includegraphics[width=0.9\textwidth]{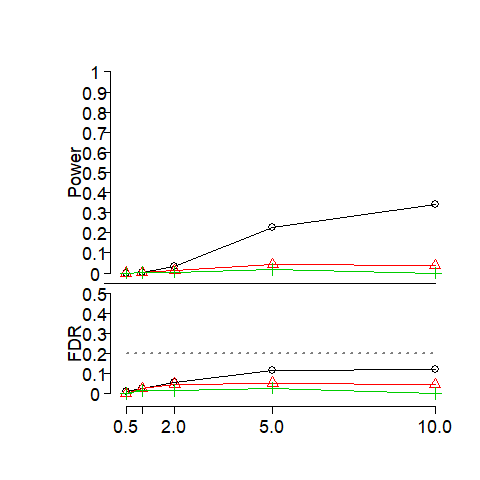}
\caption{Binary $m=10$, FDR$=0.2$}
\end{subfigure}
\begin{subfigure}[b]{.45\linewidth} 
\includegraphics[width=0.9\textwidth]{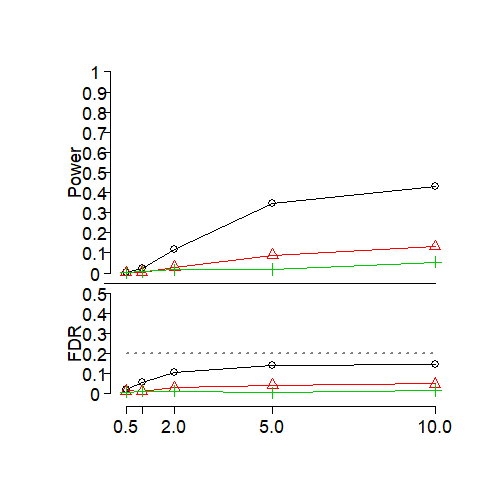}
\caption{Binary $m=20$, FDR$=0.2$}
\end{subfigure}
\caption{Simulation performance for setting 1}
\label{simuset1}
\end{figure}

\begin{figure}[h]
\begin{subfigure}[b]{.45\linewidth} 
\includegraphics[width=0.9\textwidth]{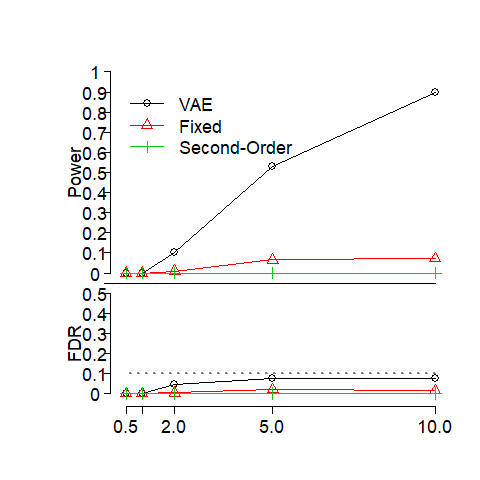}
\caption{Gaussian $m=10$, FDR$=0.1$}
\end{subfigure}
\begin{subfigure}[b]{.45\linewidth} 
\includegraphics[width=0.9\textwidth]{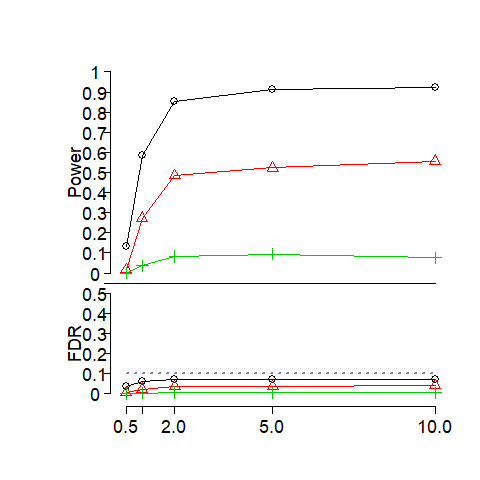}
\caption{Gaussian $m=20$, FDR$=0.1$}
\end{subfigure}\\
\begin{subfigure}[b]{.45\linewidth} 
\includegraphics[width=0.9\textwidth]{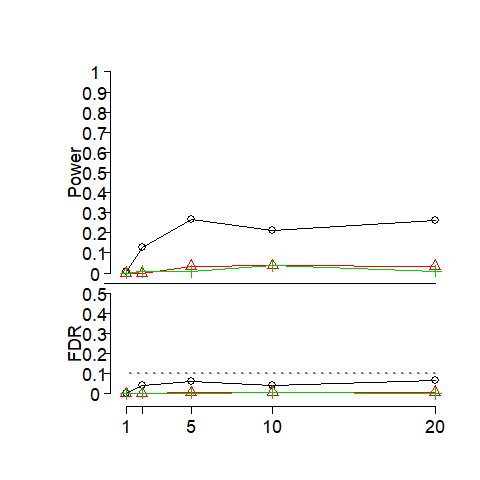}
\caption{Binary $m=10$, FDR$=0.1$}
\end{subfigure}
\begin{subfigure}[b]{.45\linewidth} 
\includegraphics[width=0.9\textwidth]{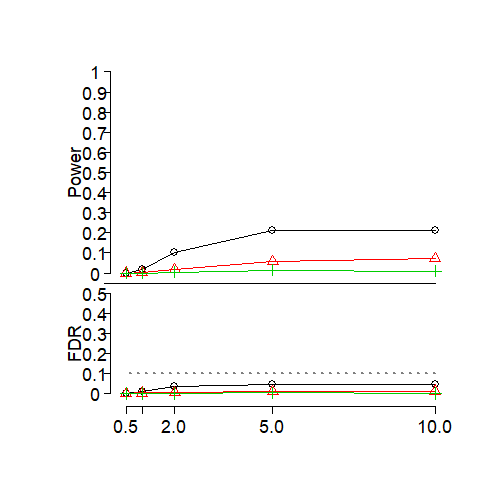}
\caption{Binary $m=20$, FDR$=0.1$}
\end{subfigure}
\caption{Simulation performance for setting 2}
\label{simuset2}
\end{figure}

In the simulation, we compare three knockoff generation schemes: a) our proposed method implemented through VAE b) the fixed knockoff generation method in \cite{barber2015} c) the second order matching in \cite{candes2018}. 
We demonstrate the performance of these methods in two simulation scenarios. In both scenarios, the sample size is 200, and number of potential predictors is 100, where the first $m$ variables are the true signals. The coefficients $\beta$ for the true predictors are alternating $\rho$ and $-\rho$, where $\rho$ is the magnitude of the signal. In Figures ~\ref{simuset1}\ref{simuset2}, the power and FDR are drawn as curves with respect to $\rho$. Results based on $100$ replications are presented for $m=10, 20$, Gaussian  and binary outcomes separately. In the Gaussian case, the error term is standard normally distributed. In the logistic regression case, binary outcome follows Bernoulli distribution with probabilities $1/(1+\exp(-X\beta))$.\\
{\bf Setting 1.} The first setting generates continuous predictors: a) Generate independent uniform $(0,1)$ distributed $200$ by $200$ random matrix. b) Let C be the Cholesky decomposition of the correlation matrix with $0.1$ on the off diagonal entries and compute $Z=UC$. c) Even column $X_{2i}$ is generated as $Z_{4i}+0.5*Z_{4i+1}^3$, and odd column $X_{2i+1}$ is generated as $Z_{4i+2}-0.5 Z_{4i+2}^2+0.5\exp (Z_{4i+3})$, where $i=0,\dots, 49$ following the python array indexing. d) Rescale $X$'s to be within range $[0,1]$ by subtracting the column $\min$ and divided by column $\max-\min$. This generation scheme is an arbitrary representation of a set of lightly correlated non-normal distributed continuous variables. Empirically, the pairwise correlation $X$ ranges from $-0.3$ to $0.6$ and the absolute value of the correlation has mean 0.07. 
The results of this setting are presented in Figure \ref{simuset1}. The signal to noise ratios of the Gaussian case approximately has the range of $0.25-0.3 \rho^2$ for $m=10$ and $0.7-0.85 \rho^2$ for setting with $m=20$.  We present the results controlled for $FDR=0.1$, because of the page limit, the $\textrm{FDR}=0.1$ binary case has low power ( although the comparison of methods are the same), thus we present results of binary outcome for $\textrm{FDR}=0.2$.
{\bf Setting 2.} The second setting generates categorical predictors:a) Generate independent standard Gaussian distributed $200$ by $100$ random matrix. b) Let C be the Cholesky decomposition of the correlation matrix with $0.1$ on the off diagonal entries. Let $Z=UC$. c) $X$ is the categorized $Z$ matrix, where $X_{ij}=\mathbf{1} (Z_{ij}>0)$. Approximately, the signal to noise ratio for the Gaussian case is $2\rho^2$ for $m=10$ and $4 \rho^2$ for $m=20$. We present results for FDR controlled at $0.1$.

{\bf Results} According to Figures \ref{simuset1} \ref{simuset2}, our proposed method implemented through VAE has FDR controlled below the predetermined threshold, and it demonstrates higher power than its two competitors. Since the data were not Gaussian, the second-order matching method has the lowest power. The assumptions of the Fixed knockoff generations holds for the Gaussian cases, thus the power of increases to about $0.4$ for setting 1 and $0.55$ for setting 2 with large signal $\rho=10$ and with $m=20$. Cases with more true signals $m=20$ demonstrate better power for Gaussian outcomes, since we control for the proportion of false discoveries, more errors are allowed with increasing number of true signals. In the binary case, although the Gaussian error assumption for using the fixed knockoff generation does not hold, FDR is still controlled in all three methods. However the powers are low.

{\bf Implementation details.} For both settings, we implemented our method with VAE \cite{kingma2014}, where the latent variables are multivariate normal with $300$ dimensions. The models were trained with batch size $25$ and for $20$ epochs with the default `adam' optimizer in `keras'\cite{chollet2015keras}. The architecture of VAE for setting 1 is as following, the encoder networks for $\mu_z$ and $\log(\sigma)$ have two hidden layers with $500$ and $400$ neurons with `tanh' activation and $L_2$ regularization with tuning parameter $0.2$. The activation for the output layer is `linear'. The decoder network has two hidden layers with the same specifications, followed by a batch-normalization layer and output with linear activation. The architecture of VAE for setting 2 is as following, the encoder networks has two hidden layers with $500$ and $400$ neurons with `relu' activation and $L_2$ regularization with tuning parameter $0.3$. The activation for the output layer is `linear'. The decoder is an one layer network with the `sigmoid' activation function and then threshold by $0.5$.

For all simulation settings and methods, we are using the signed max lambda statistics in \cite{barber2015} (see definition on page 2). We implemented the $L_1$ penalized linear and logistic regression with R package 'glmnet' with a finer grid of tuning parameter to break ties in $Z_j$ and $\tilde{Z_j}$. We used the more conservative Knockoff+ as in (2).

\section{Real Data Experiment}
\begin{figure}[h]
\begin{subfigure}[b]{.45\linewidth} 
\includegraphics[width=0.9\textwidth]{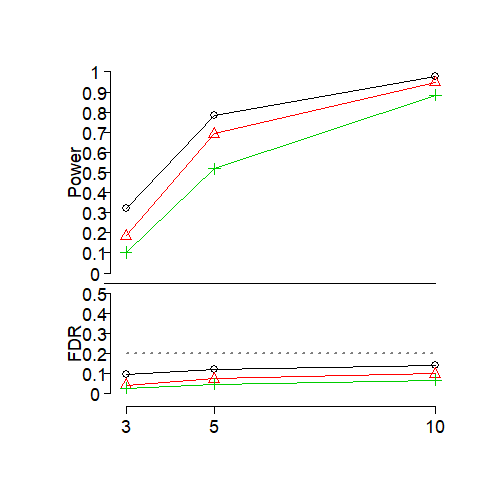}
\caption{Gaussian, FDR$=0.2$}
\end{subfigure}
\begin{subfigure}[b]{.45\linewidth} 
\includegraphics[width=0.9\textwidth]{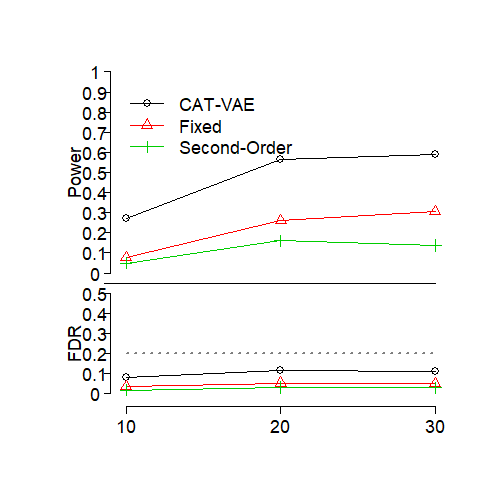}
\caption{Binary, FDR$=0.2$}
\end{subfigure}
\caption{Capture of simulated signals with real data generated  knockoffs}
\label{HIV}
\end{figure}

In the real data experiment, we demonstrate the performance of our proposed method on generating knockoffs for sparse genetic mutation data. The dataset and scientific problem is from a study aiming at detecting mutations associated with drug resistance in patients with Human Immunodeficiency Virus Type 1 (HIV-1) \cite{Rhee2006}. Due to the space limit, here we present results for the 7 protease inhibitors. There is no ground truth for which subset of mutations caused the drug resistance. Nevertheless, there is a set of $73$ treatment-selected mutations (TSMs) identified from a separate study \cite{rhee2005}, which are selected as the mutations marginally correlated with the patient treatment history with protease inhibitors. Thus to evaluate the performance of the knockoffs, we first simulates the outcome for which we know the true signal, and evaluate the FDR control and power as shown in Figure \ref{HIV}, and then we used the real data outcome and compared our selected mutations with the TSM. Notice that the TSM is not drug specific and can not be considered as ground truth, by showing number of selected mutations in TSM set we are not showing the FDR control but demonstrate reproducibility across studies. 

The normal distributed $Z$ does not fit the discrete and sparse nature of the genetic mutation data. Thus we adopted the Categorical Variation Autoencoder (CAT-VAE) \cite{eric2017,maddison2017}, where in our implementation, the latent variables $Z$ are $20$ Gumbel-Softmax distributed variable with temperature 1, each with $10$ categories; both the encoder and decoder has one hidden layer of dimension $200$. 

Figure \ref{HIV} demonstrates the results for simulated outcomes using the real data $X$. Overall speaking, our proposed method still achieves the FDR control and demonstrates the highest power among the three. The correlation is smaller than the previous simulation settings (since the X is a sparse matrix), so the other two methods has better power. Controlled for a FDR level of $0.2$, all three methods achieve a high power of 80\% very large signal at $\rho=10$ in the Gaussian case. For the binary case, our proposed method shows greater advantage in achieving more than 2-times the power of the other two methods.

Table \ref{HIVt} presents the number of selected and matched selection of mutations with the $73$ TSMs. Here we present results controlled by both Knockoff+ (as in (2)) and Knockoff (as in (1)) controlled at FDR level $0.2$.  Similar to the simulated settings, our proposed method select more mutations than the other methods, the average proportion of selected mutation that is in the TSM set for our proposed method is 75.3\%, which demonstrate reproducibility across studies. With less conservative Knockoff procedure, Fixed and Second-Order methods are able to select more mutations, where our proposed method seems be less sensitive to choice of Knockoff or Knockoff+. 

{\bf Implementation details} For these analysis, the CAT-VAE knockoffs are generated for $186$ mutations with ($\ge 4$) occurrence in the data set and the number of patients is $846$. Since each drug resistance outcome is observed in a subset of samples, when conduct analysis for each drug, we only consider mutations with $\ge 2$ occurrence. Following analysis in \cite{Rhee2006}, the outcome is a continuous variable, which is the log transformed drug susceptibility. 
For results in Figure \ref{HIV}, the linear signal is simulated as following: in each replication, randomly chose 50 predictors from 186 mutations , and compute $\mu$  as sum of $\rho x_j$'s with alternating signs. The Gaussian outcome is $\mu$ plus a standard normal noise and the binomial outcome is generated from $\textrm{Bernoulli}(\frac{1}{1+\exp(-\mu_i)})$.

\begin{table}
\caption{Number of TRM mutations/ total number of selected mutations controlled for FDR at 0.2}
\centering
\begin{tabular}{r|rlllllll}
  \hline
 && APV & ATV & IDV & LPV & NFV & RTV & SQV \\ 

&Sample Size & 767 & 328 & 825 & 515 & 842 & 793 & 824 \\
\hline
Knockoff+&  CAT-VAE & 28/28 & 8/31 & 34/49 & 27/37 & 42/54 & 38/48 & 39/52 \\ 
  &Fixed & 25/25 & 2/10 & 0 & 0 & 0 & 35/44 & 23/25 \\ 
  &Second-Order & 0 & 0 & 0 & 0 & 5/5 & 5/5 & 0 \\ 
  \hline
Knockoff& CAT-VAE & 32/40 & 9/38 & 34/49 & 28/38 & 42/54 & 38/48 & 38/50 \\ 
  &Fixed & 30/35 & 0 & 14/14 & 0 & 26/26 & 34/43 & 25/27 \\ 
  &Second-Order & 24/24 & 4/11 & 4/4 & 17/24 & 29/29 & 35/42 & 17/17 \\ 
   \hline
\end{tabular}
\label{HIVt}
\end{table}
\section{Discussion and Future Directions}
The connection between representative learning and Model-X FDR control framework has two directions. One direction is that the VAE algorithms provide a way to efficiently implement the knock generating methods proposed in this paper. The other direction is that the Model-X can also enhance the interpretability of representative learning by equipping it with the ability for finite sample FDR controlled feature selection, see example in \cite{gimenez2018}. With the natural connection between our proposed algorithm and the deep generative model for images, it is a promising future direction to investigate how to construct feature statistics to interpret image features.

The proposed knockoff generating method opens a venue for the development of new algorithms. For example, the VAE and CAT-VAE algorithms both assume $Z|X$ are independent, this can be relaxed since our method does not assume this. Some minor changes also worth further investigation, for example, instead of implementing the existing VAE algorithm with infinitesimal $q_\epsilon$, we can add a small noise to $\hat{f}(\tilde{Z})$ for generating $\tilde{X}$, with predetermined or estimated variance.

For real data applications, the existing VAE algorithms has already offered a large class of working knockoff generation models to choose from. An open question is how to evaluate the goodness of knockoffs and pick a better model. One evaluation approach is from the performance in FDR control and power in simulated settings. To this end, one can assume a Bayesian prior for the signals patterns of the data, for each knockoff model estimate the FDR and power for a class of signal patterns from several simulated settings. 
 Another approach is to evaluate the knockoff directly. This is a largely unsolved problem. An insight from Theorem 2.2 is the FDR is controlled when $q_\epsilon$ and the density of $X-\hat{f}(Z)$ are close. Therefore it is more desirable that $X-\hat{f}(\tilde{Z})$ is element-wise independent.

To sum up, the method we proposed in the paper provides a practical method for model free knockoff generation. It will stimulate broad further research and software development towards a general applicable tool for the scientific community to identify important factors with rigorous control for false discovery.


\bibliography{star}

\begin{thebibliography}{13}
\providecommand{\natexlab}[1]{#1}
\providecommand{\url}[1]{\texttt{#1}}
\expandafter\ifx\csname urlstyle\endcsname\relax
  \providecommand{\doi}[1]{doi: #1}\else
  \providecommand{\doi}{doi: \begingroup \urlstyle{rm}\Url}\fi

\bibitem[Barber \& Cand\`{e}s(2015)Barber and Cand\`{e}s]{barber2015}
Rina~Foygel Barber and Emmanuel~J. Cand\`{e}s.
\newblock Controlling the false discovery rate via knockoffs.
\newblock \emph{Ann. Statist.}, 43\penalty0 (5):\penalty0 2055--2085, 10 2015.

\bibitem[Barber et~al.(2018)Barber, Cand\`{e}s, and Samworth]{barber2018}
Rina~Foygel Barber, Emmanuel~J. Cand\`{e}s, and Richard~J. Samworth.
\newblock Robust inference with knockoffs.
\newblock \emph{{arXiv:1801.03896 [stat.ME]}}, 2018.

\bibitem[Benjamini \& Hochberg(1995)Benjamini and Hochberg]{BH1995}
Yoav Benjamini and Yosef Hochberg.
\newblock Controlling the false discovery rate: A practical and powerful
  approach to multiple testing.
\newblock \emph{Journal of the Royal Statistical Society. Series B
  (Methodological)}, 57\penalty0 (1):\penalty0 289--300, 1995.

\bibitem[Cand\`{e}s et~al.(2018)Cand\`{e}s, Fan, Janson, and Lv]{candes2018}
Emmanuel Cand\`{e}s, Yingying Fan, Lucas Janson, and Jinchi Lv.
\newblock Panning for gold: ‘model-x’ knockoffs for high dimensional
  controlled variable selection.
\newblock \emph{Journal of the Royal Statistical Society: Series B (Statistical
  Methodology)}, 80\penalty0 (3):\penalty0 551--577, 2018.

\bibitem[Chollet et~al.(2015)]{chollet2015keras}
Fran\c{c}ois Chollet et~al.
\newblock Keras.
\newblock \url{https://keras.io}, 2015.

\bibitem[Gimenez et~al.(2018)Gimenez, Ghorbani, and Zou]{gimenez2018}
Jaime~Roquero Gimenez, Amirata Ghorbani, and James Zou.
\newblock Knockoffs for the mass: new feature importance statistics with false
  discovery guarantees.
\newblock \emph{{arXiv:1807.06214 [stat.ML]}}, 2018.

\bibitem[Jang et~al.(2017)Jang, Gu, and Poole]{eric2017}
Eric Jang, Shixiang Gu, and Ben Poole.
\newblock Categorical reparameterization with gumbel-softmax.
\newblock \emph{ICLR}, 2017.

\bibitem[Kingma \& Welling(2014)Kingma and Welling]{kingma2014}
Diederik~P. Kingma and Max Welling.
\newblock Auto-encoding variational bayes.
\newblock \emph{ICLR}, 2014.

\bibitem[Maddison et~al.(2017)Maddison, Mnih, and Teh]{maddison2017}
Chris~J. Maddison, Andriy Mnih, and Yee~Whye Teh.
\newblock The concrete distribution: A continuous relaxation of discrete random
  variables.
\newblock \emph{ICLR}, 2017.

\bibitem[Rezende et~al.(2014)Rezende, Mohamed, and Wierstra]{rezende2014}
Danilo~Jimenez Rezende, Shakir Mohamed, and Daan Wierstra.
\newblock Stochastic backpropagation and approximate inference in deep
  generative models.
\newblock 2014.

\bibitem[Rhee et~al.(2005)Rhee, Fessel, Zolopa, Hurley, Liu, Taylor, Nguyen,
  Slome, Klein, Horberg, Flamm, Follansbee, Schapiro, and Shafer]{rhee2005}
Soo-Yon Rhee, W.~Jeffrey Fessel, Andrew~R. Zolopa, Leo Hurley, Tommy Liu,
  Jonathan Taylor, Dong~Phuong Nguyen, Sally Slome, Daniel Klein, Michael
  Horberg, Jason Flamm, Stephen Follansbee, Jonathan~M. Schapiro, and Robert~W.
  Shafer.
\newblock Hiv-1 protease and reverse-transcriptase mutations: Correlations with
  antiretroviral therapy in subtype b isolates and implications for
  drug-resistance surveillance.
\newblock \emph{The Journal of Infectious Diseases}, 192\penalty0 (3):\penalty0
  456--465, 2005.

\bibitem[Rhee et~al.(2006)Rhee, Taylor, Wadhera, Ben-Hur, Brutlag, and
  Shafer]{Rhee2006}
Soo-Yon Rhee, Jonathan Taylor, Gauhar Wadhera, Asa Ben-Hur, Douglas~L. Brutlag,
  and Robert~W. Shafer.
\newblock Genotypic predictors of human immunodeficiency virus type 1 drug
  resistance.
\newblock \emph{Proceedings of the National Academy of Sciences}, 103\penalty0
  (46):\penalty0 17355--17360, 2006.

\bibitem[Sesia et~al.(2018)Sesia, Sabatti, and Cand\`{e}s]{sesia2018}
M~Sesia, C~Sabatti, and E~J Cand\`{e}s.
\newblock Gene hunting with hidden markov model knockoffs.
\newblock \emph{Biometrika}, 2018.

\end{thebibliography}
\bibliographystyle{iclr2019_conference}

\end{document}